# Adult-child pairs walking down stairs: Empirical analysis and optimal-step-based modeling of a complex pedestrian flow, with an exploration of flow-improvement strategies

Chuan-Zhi Xie[ab], Tie-Qiao Tang[a*], Bo-Tao Zhang[c], Alexandre Nicolas[b*]

a) School of Transportation Science and Engineering, Beijing Key Laboratory for Cooperative Vehicle Infrastructure Systems and Safety Control, Beihang University, Beijing 100191, China

b) Institut Lumière Matière, CNRS & Université Claude Bernard Lyon 1 & Université de Lyon, 69622, Villeurbanne, France

c) Department of Architecture and Civil Engineering, City University of Hong Kong, Kowloon Tong, Hong Kong Special Administrative Region, China

**Abstract**: Pedestrian egress from training schools in the after-class period (especially in China, as children walk down stairs together with their parents) raises practical concerns related to degraded flow conditions and possible safety hazards, but also represents a challenge to mainstream modeling approaches for several reasons: they involve heterogeneous groups (adult-child pairs), which are hardly studied compared to their homogeneous counterparts, in a complex geometry, made of staircases connected by a platform where pedestrians rotate, and over a wide range of densities. In light of our field observations at a training school in China, we develop a semi-continuous model which quantitatively reproduces the collective dynamics observed

___________________________

*Corresponding author:

tieqiaotang@buaa.edu.cn (T.-Q. Tang), alexandre.nicolas@polytechnique.edu (A. Nicolas)

empirically and enables us to assess some guidance strategies to improve egress efficiency. In this model, which extends the optimal step approach, adults and children are described as ellipses with prescribed relative positions that evolve by successive steps handled by an event-driven algorithm, along a spontaneous semi-circular path on the platform, but may deviate from it in crowded conditions by selecting their preferred next positions. In line with the observations, most pairs walk side by side overall and only a small fraction sometimes make a detour to avoid queuing. Turning to the guidance strategies, we find that promoting front-back pairing may increase the flow, while urging people to make more detours may be counterproductive. Perhaps even more relevantly, the intuitive measure consisting in desynchronizing the flows from the different floors by shifting the ends of classes succeeds in reducing the egress time by at least 10%.

## 1. Introduction

With ever growing urbanization, the number of indoor gathering sites has constantly increased, and is expected to keep growing. The crowded settings in these facilities may degrade the pedestrian flow conditions and possibly raise safety hazards, caused by pedestrian movement factors (internal) and architectural design (external causes). Schools are an example of such venues where crowded conditions may become perilous: for instance, in 2017, in a primary school in Henan Province, China, a student died and 21 students were injured in the after-class period due to a stampede caused by chaos in the crowd [1]. Key in the effort to curb these hazards is a detailed understanding of pedestrian motion and how pedestrians' properties (e.g., gender, age) affect their motion in some specific scenarios has further been paid attention. Currently, researchers mostly resort to two approaches to this end: experiments, whereby real pedestrian behavior characteristics can be extracted, and simulations, which enable researchers to collect and replicate efficiently and explore some scenarios that are hardly amenable to experimental studies. For example, by classifying people by age (i.e., children, adults and the elderly, etc.), researchers conducted empirical studies aiming to lay a reliable experimental-based resource for proposing, validating, and evaluating the corresponding pedestrian flow models [2-5]; the variations of pedestrian behavior depending on the considered setup have also been explored, with focuses on bottlenecks, corridors, vertical walking facilities (i.e., stairs and escalators), classrooms, merging area, etc. [6-8]. Besides, the pedestrian grouping and following behaviors have also attracted considerable attention [9, 10]. All these endeavors certainly deepen our understanding of pedestrian motion and may lead to better emergency or non-emergency measures for pedestrian guidance.

However, among the practically relevant situations, the case of mixed adult-child pedestrian flow still partly lies in the shade, despite its pervasiveness. One survey conducted by the Chinese Ministry of Education in 2018 mentions that among 128418 investigated training schools (i.e., typical places featuring adult-child pairs), 7.83% of them (10051 in total) have significant safety hazards [11]. Bearing these risks in mind, why does this type of crowd require singular attention, eluding the capacities of current modeling tools? At least three reasons contribute to this specific status: (1) adult-child pairs are specific social groups, whose members are intimately connected both mentally and physically, and possibly holding hands; (2) the pedestrian attributes within each pair are very heterogeneous, with different spontaneous step lengths, walking speeds, etc. between the adult and the child; (3) compared to pedestrian traffic (flow) under road environment, the indoor venues in which such pairs are encountered typically have a more complex spatial arrangement, such as multi-floor sites (note that staircases present even more safety hazards for pedestrian movement than floors), with a density that can reach relatively high values in some specific periods. These reasons call for particular caution when applying generic commercial software or academic models to adult-child mixed flows, with the prospect of assessing guidance strategies to improve their operational efficiency and/or safety.

Accordingly, in this study, we first conduct a careful empirical analysis of field observations at a Chinese training school (similar to Ref. [12], see Fig. 1), in Section 2. These observations then serve as a solid basis for the development of a microscopic adult-child mixed pedestrian model, which extends the optimal-step approach (Section 3). Section 4 is dedicated to the quantitative comparison of the model predictions and the empirical results. Finally, in Section 5, some management strategies and guidance measures are implemented and tested numerically.

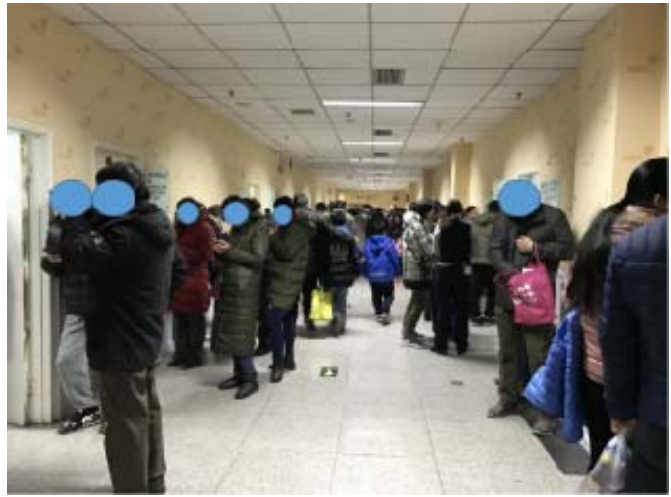

Fig. 1 Photograph of a typical adult-child crowd at Lily English training school, Beijing, taken on 03/01/2018 in the after-class period, when adults were picking up their children.

## 2. Empirical study

Empirical studies are instrumental in understanding the way in which pedestrian flows self-organize in fairly complex settings such as those under study and extracting both qualitative and quantitative information (e.g., velocity, grouping pattern, etc.) about the structure of the flow. These data can then help to build and calibrate pedestrian movement models.

## 2.1 Literature review

Experimental studies about pedestrian flow are generally divided into two main categories: controlled experiments (in which participants move within an established experimental framework) and field observations (where real-life pedestrian movements are passively observed and recorded). Since this paper mainly focuses on adult-child pairs' movement on staircases, we used the following two keywords for our review of previous empirical works: pair **or** staircase.

The walking pattern of groups during egress or evacuation processes was studied by Fu et al. [13] as well as Moussaïd et al. [14]; Fu et al. point out that pairs of pedestrians (i.e., groups of size 2) tend to walk abreast (side-by-side) whereas triads (groups of size 3) tend to display a V-shape, with the central pedestrian slightly backwards; Moussaïd et al. report the same walking pattern for triads and find, in addition, that as the density increases the walking pattern of triads will transform from side by side way to a more pronounced V-shape. Meanwhile, the influence of pedestrian groups on walking speed and evacuation efficiency has also attracted attention; Gorrini et al. [15] focus on crowds of both the elderly and adults, corroborating the negative impact of social interactions within groups on speed; similarly, Wei et al. [16] also find that the grouping behavior negatively impacts the walking speed, as well as the step frequency of a particular pedestrian. Contrary to Refs. [15] and [16], Von

Krüchten and Schadschneider [17] claim that from the perspective of overall evacuation efficiency, due to self-ordering effects, grouping behavior may play a positive role in promoting the overall evacuation process. At this stage, let us point out that even though Refs. [15-17] consider different populations (i.e., students, adults, and the elderly) and adopt different indices to explore the impact of groups, all are mainly concerned with fairly homogeneous groups of people. To what extent their conclusions are altered in the presence of heteroclite pairs, still is to be explored.

An additional source of complexity of the scenario under study is the staircase, which is typically the primary egress component for buildings with distinct hazardous features [18]. Indeed, the pedestrians' behavior on staircases is way more complicated and less explored than on a flat floor [13, 19]. Controlled experiments are frequently adopted to extract the essential attributes of staircase pedestrian flow; for example, Chen et al. [20] performed a series of single-file experiments on staircases (both ascending and descending) to explore the fundamental diagram and the inner relation between variables (e.g., speed-headway relationship). Some researchers further investigated how pedestrian attributes (e.g., age, gender, and occupation) affect the egress process [21]. Specifically, Kholshchevnikov et al. [22] use preschool as the scenario with different age groups of children to record objects' behavior and summarize a general formula to describe the speed-density under such experimental conditions. Also, the external factors that affect staircase emergency evacuation have been explored; for example, firefighters may need to ascend to rescue while the evacuating crowd goes down [23].

To sum up, the above studies seldom consider heterogeneous pedestrian grouping or heterogeneous pedestrian flow on staircases. In fact, given the differences between adult-child pairing (or grouping) and homogeneous (adult or child) pairing, it remains unclear how adult-child mixed flow will operate on staircases, underlining a research gap: in light of the ubiquity of buildings serving for adult-child mixed flows, how to guarantee the safety and efficiency for adult-child mixed flow on staircases? This paper carries out an empirical study to find out the pair's movement mechanism to partly contribute to this question.

2.2 Research scenario

To obtain the movement characteristics of adult-child mixed pedestrian flow on staircases, we conducted a field study[1] at a Chinese training school. The settings are composed of three parts: the first part of the staircase (denoted as stair-1 in short), the second part of the staircase (denoted as stair-2 in short), and the flat platform connecting stair-1 to

---

[1] The empirical study was examined jointly by the head director of School of Transportation Science and Engineering, Beihang University, and the manager of Xueersi Training School, Beijing, and it received their approval.

stair-2 (denoted as platform in short). In the after-class period, pedestrians use these facilities to egress from the building. The geometry of the staircase is sketched in Fig. 2, where one may notice, i.e., i) the width, length, and height of each step of the staircase are 1.3-m, 0.3-m, 0.15-m, respectively; ii) each flight of stairs consists of 12 steps; iii) the width and length of the platform are 2.7-m and 2-m, respectively. Regarding the platform (in blue on Fig. 2), it can be accessed from the stairs (stair-1), but also via a central gate of width equaling 1.3-m. Besides, two rubbish bins are located on both sides of the gate, with 0.4-m length and 1-m width.

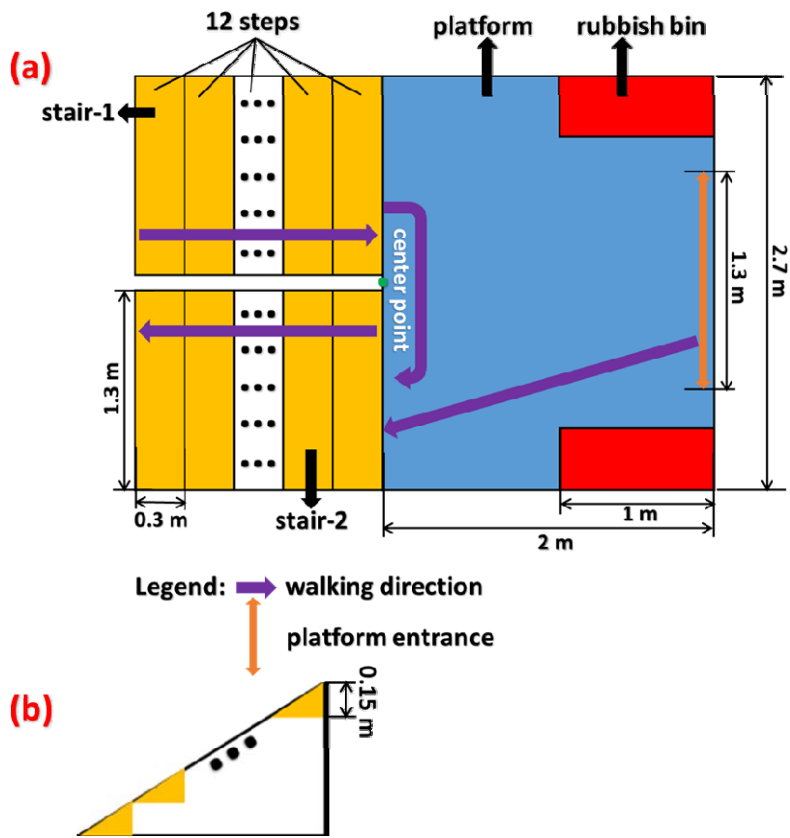

Fig. 2 Schematic view of the research area, where (a) is the top view of the whole area, and (b) is the side view of the staircase.

2.3 Empirical analysis

To film the egress in the after-class period, two HD surveillance cameras[2] were fixed to the wall and could be adjusted $360°$ to guarantee the best viewing angle (see Fig. 3(a)). One of the two cameras recorded the pedestrian movement data on stair-1 (see Fig. 3(b)), and the other recorded platform and stair-2 (see Fig. 3(c)). Quantitative data were extracted manually from the video, mostly through visual inspection; these data included the entrance time and number of pedestrians in different areas, their group formations, the number of steps on the platform, the way the parent or the child adjusted their speed, among others.

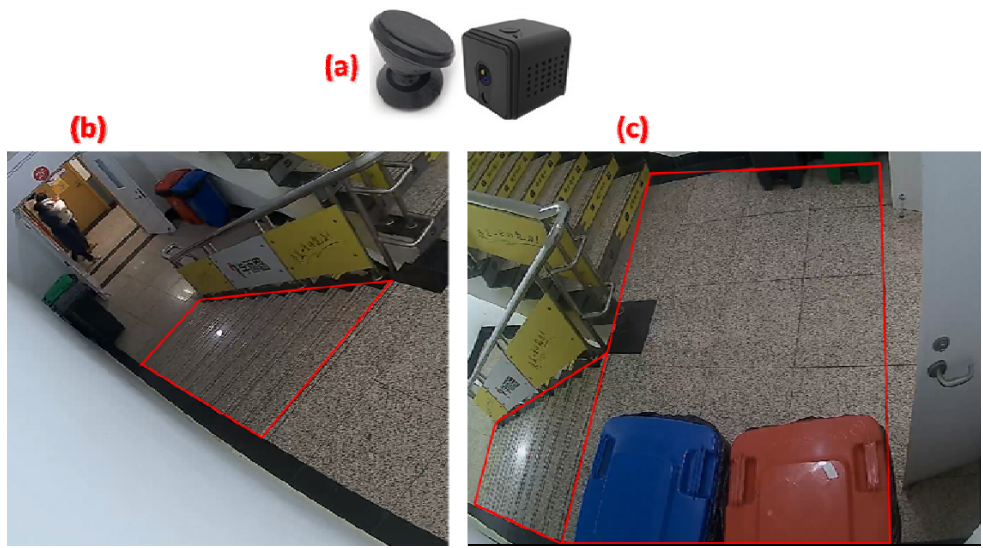

Fig. 3 (a) Photograph of the camera; (b) view of stair-1, delimited by red lines, as filmed by one camera; (c) view of the platform and stair-2, delimited by red lines, as filmed by another camera.

---

[2] Camera details: brand: JooAn; resolution: 720P; fps: 20.

The observation period, from 20:19:46 p.m. to 20:28:10 p.m. on 01/01/2021 (i.e., 505s overall), covers the evening after-class period of the training school. In this time interval, 222 pedestrians entered the observation area from the entrance of stair-1, among whom 94 adult-child pairs; 108 pedestrians entered the platform from platform-gate, of which 47 adult-child pairs. One of the key focuses of this study is the pairing behavior of pedestrians, and the physical formations among pedestrians in grouping (pairing) states differ and may influence the movement of other pedestrians [14, 17]. Careful observation of the video led to the definition of three formations of adult-child pairs : left-right pairing with adult-outer-child-inner (denoted as **child:inner**), left-right paring with adult-inner-child-outer[3] (denoted as **child:outer**), and child-front-adult-back pairing (denoted as **child:front**) (see Fig. 4). The first two pairing modes are dominant, representing approximately 90% of all pairs. Interestingly, the pairing modes of adult-child pairs markedly differ from those of ordinary dyads (e.g., compared to Ref. [24] conducted by Costa). While members of an 'ordinary' dyad will typically walk with little physical contact or interaction, thus giving a large variety of pairing modes, adult-child pairs are structured by the adults' willingness to control and protect their children; for example, the **child:front** pairing mode gives each adult a stronger control over his/her child, much stronger than had the child been behind the adult.

---

[3] In this study, pedestrian flows from stair-1 and platform-gate are likely to encounter certain competition and conflict. Therefore, we define the side where the two pedestrian flows may conflict as outer and the other side as inner. That is, based on the pedestrian pair's face orientation, for the pedestrian pair from stair-1, the left side is regarded as outer and the right side is regarded as inner; for the pedestrian pair from platform-gate, the left side is regarded as inner and the right side is regarded as outer.

On the basis of our careful observations, we can make the following empirical assertions:

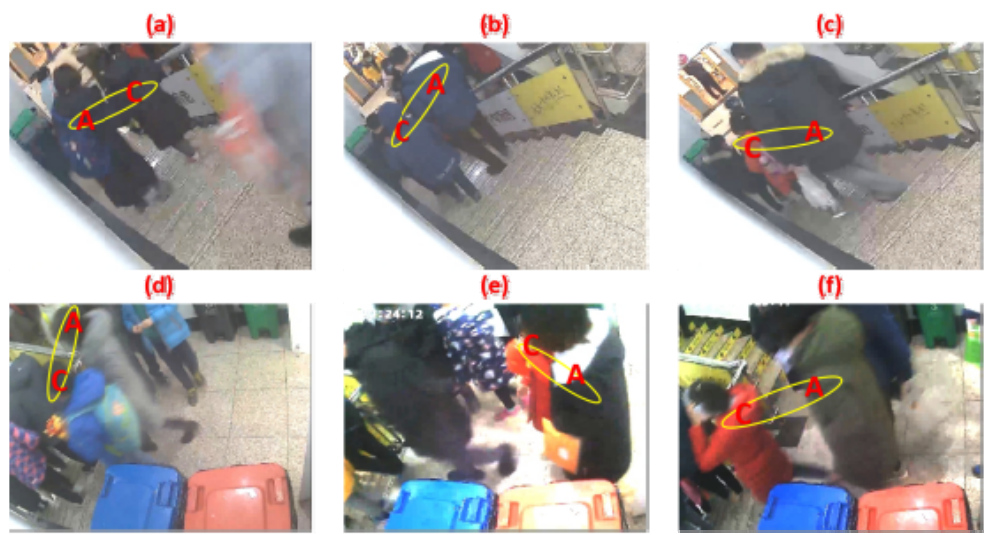

Fig. 4 Adult (A)-child (C) pairing modes, where (a)-(c) represent **child: inner**, **child: outer**, and **child:front** (on stair-1), respectively; (d)-(f) reflect **child:inner**, **child:outer**, and **child:front** (coming from the gate), respectively.

(1) Pairs tend to maintain their pairing modes during their whole passage through the recording area.

(2) On the staircase, pair members move at most one step down at a time.

(3) In contrast, on the platform, they follow a semi-circular path centered around 'center point' (see Fig. 2), consistently with the earlier report of Hyun-seung et al. (see Fig. 3 in Ref [25]), and make different numbers of steps to cross the platform, ranging from 4 to 7 (for pairs coming from stair-1), owing to the different pairing modes and initial landing position on the platform.

(4) On the platform, for adult-child pairs coming from stair-1, the outer pedestrian controls the walking speed of the pair (i.e., the outer pedestrian is the 'leader'). Accordingly, an adult-child pair 'led' by an adult has a longer step-length (approximately 0.5-m under free flow condition) than its counterpart led by a child (approximately 0.45-m).

(5) With increasing density on the platform, adult-child pairs make more steps to cross the area: a pair from stair-1 makes as little as 5 steps in free flow and up to 8 steps in dense conditions. In our case, the density fluctuated considerably and peaked at a maximum value corresponding to 32 people simultaneously in the global area made of the two staircases and the platform (approximate density on the platform: 2.17 ped/m²). This peak is caused by the fact that the majority of classes end around the same time (see Fig. 10).

(6) In the case of congestion or constrained motion due to conflicts, most adult-child pairs will wait and stick to their original semi-circular paths. Nevertheless, a few pairs will instead make a detour to avoid congestion (i.e., to find available positions closer to the destination).

(7) For two pedestrian flows entering platform from stair-1 and platform-gate, oscillations between the two flows are observed at the confluence.

Albeit interesting *per se*, these observations hardly suffice to gauge the efficiency of the collective egress flow or explore the effect of variations in the settings or pairing behaviors. Therefore, we turn to the development of a model premised on the empirical findings, but with broader applicability.

## 3. Extended Optimal Step Model for adult-child mixed pedestrian flow

Here, we aim for a robust, computationally efficient model which can both reproduce our empirical observations at least semi-quantitatively and rest on rules that are general enough to make it applicable to a wider range of situations (e.g., variations of the scenario of study), rather than ad hoc rules.

3.1 Survey of existing models

In our search for a suitable modeling framework, we first noticed that none had been developed for the particular (but practically relevant) topic of mixed adult-child pedestrian flows on staircases. Let us then broaden the focus and consider other kinds of groups. In that broader context, among the few models studying the motion of pairs on stairs, Li et al. [26] developed a cellular automaton (CA) to simulate the flow of groups of 2 or 3 pedestrians at a light-rail transfer station, where people have the choice between an escalator and a stairway. This choice was entrusted to a support vector machine and a group leader was set responsible for the group decisions; nevertheless, the motion of people on the stairs was not probed in detail.

That being said, inspiration could be drawn from previous pedestrian models focusing either on dyads **or** on staircases. Pedestrian motion in our scenario can be split into two phases: on the stairs and on the inter-stair platform. The discrete nature of the stairs naturally favors a discrete approach (see for instance Refs. [27-29]), which has led researchers to use the same discrete approach for the connecting platform as well, for internal consistency. This choice may however be questioned, as the pedestrian body rotates on the platform to accommodate the swift change in the direction of motion. In a discrete approach, this directional change will occur all of a sudden, whereas the empirical study of Hyun-seung et al. [25] (as well as ours, see above) rather points to a semi-circular trajectory. Bao and Huo [30] underline the very same deficiencies of existing models (which they overcome by bringing in a more continuous treatment compared to previous studies).

The Optimal Step Model (OSM for short) [31, 32], propounded by Seitz and Köster, seems to offer a good compromise between the realism of continuous space and the simplicity of CA. OSM is indeed able to describe steps in arbitrary directions (only limited by the optimization algorithm and numerical resolution): at each time step, each agent chooses their next position on a circle of radius equal to their step length (see Fig.1 of Ref. [31]). Update times are therefore discrete and given by the step duration. Thus, OSM benefits from the simplicity, robustness, and computational efficiency of CA and the spatial accuracy of continuous models (which update the pedestrian's movement with very short time steps, e.g., $\Delta t = 0.01\text{s}$). A very recent work of Bao and Huo [30] (conducted independently of ours) provides a solution of depicting **single agents'** translational and rotational behavior by introducing a total potential (i.e., which reflects the pedestrian-pedestrian and pedestrian-environment interaction) into the original OSM considering structural characteristics of the connection part of the staircase.

Nevertheless, it is not suitable for our endeavors in its native version. Indeed, first, it includes no pairing strategies. In this respect, inspiration could be drawn from Zanlungo et al. [33], who reproduced the dynamics of the relative motion within dyads and triads under dilute conditions by introducing a pairing (or grouping) potential. At higher density, Han and Liu [34] modified the social force model with an information transmission mechanism, especially between groups members. These two references demonstrate that group behavior can be implemented in continuous models, with good performance and scalability. On the other hand, the strategies used in CA are often less flexible, but also easier to implement and much less prone to giving rise to spurious behaviors. In this context, Crociani et al. [35] managed to model cohesive, but not strictly rigid groups by introducing a utility function into a CA model. Focusing on the child crowd, Chen et al. [36] managed to model the behavior of groups of children during the non-emergency egress from a classroom with an extended CA, which reached a relatively good performance compared to their empirical data in a fairly simple setting, where desks and chairs structure the classroom space. One of us recently gave a more comprehensive summary of the modeling options for ordinary pedestrian groups [37]. Yet, these options largely ignore the specifics of adult-child pairs, who admittedly display some similarities with more homogeneous groups with a leader-follower structure [38, 39], but are also more compact and more tightly bound, especially when they hold hands. Modeling methods should thus be adapted accordingly.

Secondly, the OSM does not specifically address the issues that arise at high densities on the platform, i.e., how the actual shapes of the agents will matter (note that Seitz and Köster do mention the extension capability of changing agents' shape instead of circles, but leave the implementation remains open [31]).

3.2 Introduction of an Extended OSM (EOSM)

Here, we substantially extend the OSM for our purposes, in the following directions: i) each pedestrian is described as an ellipse; ii) adult-child pairs are introduced, with a rigid structure; iii) small steps are made possible at high densities accounting for agents in pair, by allowing pairs an alternative to walking the maximum step length or halting: they can choose to make a smaller step, of length between zero and maximum step length (note that under the single pedestrian crowd context, the 'smaller step' concept has also been afterwards raised in Ref. [40]).

Overall, the rules of our EOSM mostly derive from the following general principles, along with the empirical observations of Section 2:

(1) Agents cannot overlap in space.

(2) For each adult-child pair, the pairing mode is kept constant throughout the simulation.

(3) Agents go down the stairs (at most) one step at a time and intend to follow a semi-circular trajectory (intended path) upon reaching the platform.

(4) On the platform, agents will never detour (swerve from the semi-circular intended path) if their next step does not lead to a spatial conflict.

(5) Agents can make a step of variable length, between zero and a maximum step length, but the time taken to make the step (time gap between successive updates) is constant for each pair of agents, while it may differ between pairs to account for their different free walking speeds.

(6) The 'optimal' step taken by the agents is governed by the maximization of a given utility function, which is here simply the proximity to a given target, with no retreating behavior (i.e., agents can not reach positions with less utilities).

3.2.1 An event-driven algorithm with joint decisions for pairs of agents

Once the agents entered the setup, their positions will be updated asynchronously in a discrete series of moves, with the help of an event-driven algorithm which keeps track of the list of next update times for all pairs of agents. Concretely, the adult-child pair with the most imminent update time (as explained below) is selected from the list and allowed to move. In other words, competitions for future positions are solved on the basis of the order of priority in the list of update times. Should two or more pairs fortuitously share the same update time, this competition is resolved randomly, by picking one pair at random, thus making the updates fully asynchronous.

Once a pair has moved (or abstained from moving), their 'next update time' is incremented by their pair-specific time interval and re-inserted into the list of update times, and so on, until all agents have exited. The pair-specific time interval is related to the characteristics of motion by the equation: interval=step-length/walking-speed, where step-length and walking-speed refer to the step length and walking speed during *free motion*; these parameters are shared by the adult and the child in each pair (consistently with our observation of a pair motion controlled by a leader in each pair), but may differ between pairs. More specifically, on the staircase, the step length coincides with the length of a staircase step and the pair's walking speed is that of the child. When they reach the platform, these pairs change these parameters for the step length and walking speed of the adult, for the **child:inner** formation, or of the child, for the **child:outer** and **child:front** pairing modes.

3.2.2 Body shape and stepping options

Each agent is characterized by a body shape represented by an ellipse, with the body width as the long axis (0.5-m for adults, 0.4-m for children) and the body thickness as the short axis (0.2-m for both adults and children). Note that using this more realistic shape, compared to a disk, becomes a necessity at high densities, when people are tightly packed, insofar as disks of radius half the shoulder width strongly underestimate the maximal packing density of the crowd. The no-overlap condition outlined in our principle-(1) implies that contemplated steps that lead to an overlap between two ellipses (as checked by a dedicated MATLAB routine) are forbidden; the closest two ellipses can get is just in contact. Provided the no-overlap criterion is met, members of a pair will opt for an optimal step that maximizes their common utility, among a discrete set of options. For most pairs, these options are restricted to the possibility of making smaller steps than the 'free' step length on the platform. But, to reflect our observation of some rare, but non-negligible detours, some pairs are assigned with 'detouring' and 'randomization probability' attributes when making decisions, once and for all, which enable them to also vary their walking direction or halt, among a discrete set of options. In the following, we detail these rules depending on whether the agents are on the stairs, on the platform coming from the stairs, or on the platform coming from the gate. Finally, to bar the possibility (offered by our discrete updates) of 'jumping' over an agent, if an agent's step leads to a new position without any overlap, but the intermediate positions leading to the new position violate the no-overlap condition, then the move will be prohibited, even though the final position appears to be acceptable.

3.2.3 Motion on the staircase

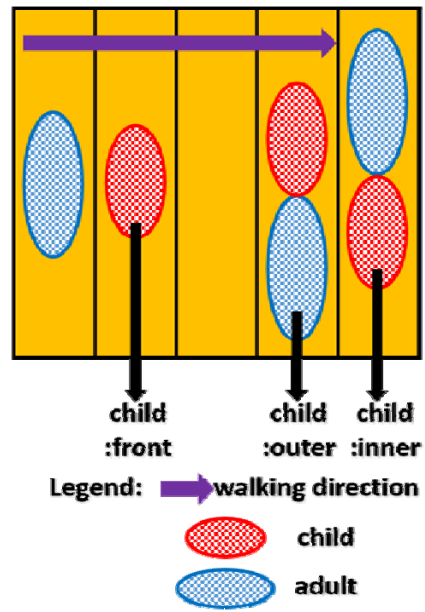

Fig. 5 Illustration of the relative positions of adult-child pair members on the staircase.

On the stairs, agents keep their body orientation fixed, with their major axes aligned with the crosswise direction. **Child:inner** and **child:outer** pairs are always on the same step, whereas **child:front** pairs stand at the same transverse position, but on successive steps, without any gap between child and adult, as illustrated in Fig. 5. When the time for their next update comes, a pair may walk down *one* step (step-length in Fig. 2(a), no step is skipped) if this does not lead to an overlap with another agent. Regarding transverse motion, pairs with no detouring attribute maintain their transverse position throughout the flight of stairs, whereas pairs featuring a detouring behavior undergo a transverse fluctuation $l_\perp$ at every step, where $l_\perp$ is randomly drawn from a normal distribution $N(0,(D\_x/3)^2)$ of standard deviation $D\_x > 0$, calculated such that $D\_x^2 + (S_1/2)^2 = S_1^2$.

3.2.4 Motion on the platform for agents coming from the stairs

Pedestrians coming from the staircase enter the platform via the 'connecting area', whose shape is a rectangle with 0.3-m as the short-side length (i.e., the length of each tread) and 1.3-m as the long-side length (i.e., the width of stair), which is still handled as a staircase step. Upon reaching this area, they start to move with the 'platform rules' exposed in the following.

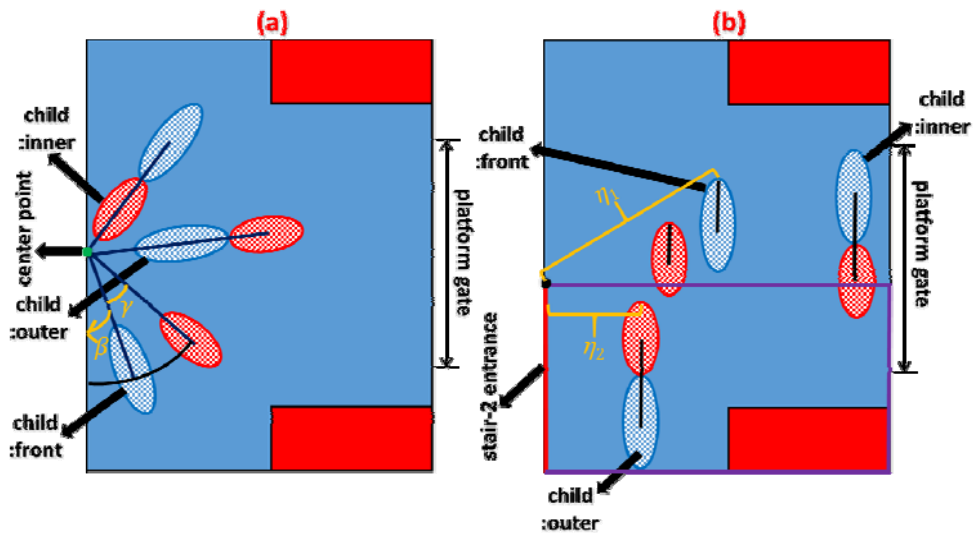

Fig. 6 Geometric sketch and notations pertaining to the motion on the platform for (a) pairs coming from the staircase and (b) pairs from the platform gate.

On the platform, agents embark on a rotating circular motion centered on 'center point' (shown as the green dot in Fig. 2(a) and in Fig.

6(a)), that is to say, **child:inner** and **child:outer** pair members remain aligned radially, with their major axes along the same spoke of the circle, whereas **child:front** pair members have their centers at the same distance from 'center point' and maintain a fixed angle $\gamma$ (see Fig. 6(a)) between them (the exact value of $\gamma$ depends on the angular difference when the two pedestrians stand together on the platform for the first time).

At every step, agents make a step of length between zero and the maximum allowed step length, i.e., following principle-(2), striving to minimize the angular distance $\beta$ to the next flight of stairs (as shown in Fig. 6(a)); $\beta$ will thus decrease from $\pi$ to 0 during the motion on the platform). More precisely, pedestrians without detouring attribute will select the longest step length possible that does not lead to an overlap, among $k_r$ equally spaced, discrete options (green dots in Fig. 7), while remaining on the same semi-circle, i.e., at the same radial distance from the center point. On the other hand, the minority of pairs featuring a detouring behavior are additionally allowed to choose their stepping direction, among $k_\theta$ equally spaced, discrete options (red dots in Fig. 7). Combined with the step-length options, this leads to a number $k_r \cdot k_\theta$ of options, represented by red and green dots in Fig. 7. The 'optimal step' which is selected is that which minimizes the new angular distance $\beta$ without leading to any collision for the child or the adult. Note that this is consistent with our general principle-(5), insofar as agents will always make an optimal step on the same semi-circle if there is no spatial conflict.

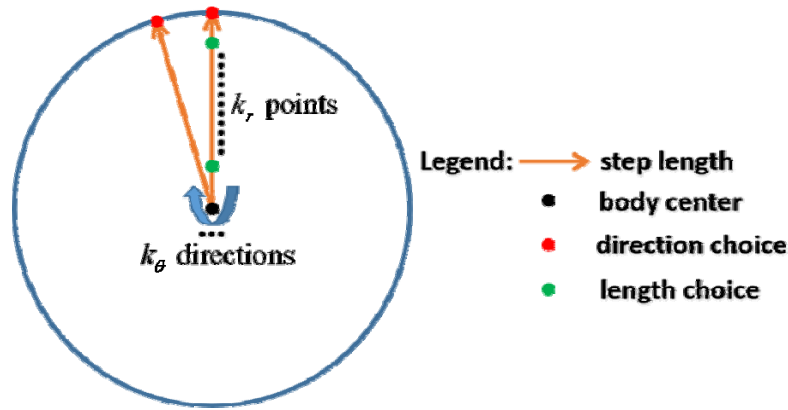

Fig. 7 Sketch illustrating the stepping options considered by an agent initially located at the center of the circle.

3.2.5 Motion on the platform for agents coming from the gate

Agents coming from the platform gate aim to minimize the distance $\eta$ to stair-2, while maintaining their pairing mode (principle-(2)). **Child:inner** and **child:outer** pair members have their major axes aligned on a line parallel to the edge of stair-2 (shown in red in Fig. 6(b)), whereas **child:front** pair members have their major axes on two lines also parallel to the staircase edge, but separated by a fixed distance (equal to the distance separating them when they both first enter the platform). Technically speaking, the distance $\eta$ is calculated as the distance between the farthest side point of the ellipse and the edge of the first step of staircase 2 (see Fig. 6(b)); this boils down to the distance $\eta_2$ between the edge and the major axis of the ellipse if the latter is fully contained in the rectangle represented in purple in Fig. 6(b), or to the distance $\eta_1$ otherwise.

It follows that pairs without detouring attribute will always choose position by trying to minimize $\eta$ for a fixed step length, shown as a dashed purple line in Fig. 8; along this direction (intended path), they will opt for the longest step length that does not lead to any overlap, among $k_r$ discrete options. As for pairs with a detouring proclivity may additionally choose their direction, among $k_\theta$ options (applying to pairs from both the staircase and gate); the 'optimal' step chosen by the two pair members is the one that minimizes $\eta$ without inducing any collision.

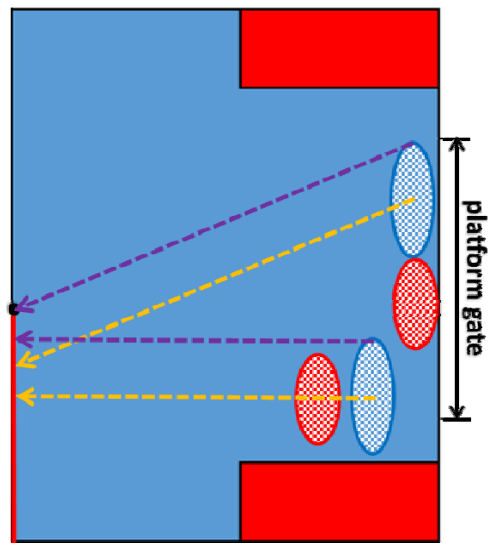

Fig. 8 Intended path calculation for pairs from platform gate.

Finally, when a pair member reaches the 'connecting area', whose shape is a rectangle with 0.15-m as the short-side length and 1.3-m as the long-side length (i.e., the width of stair-2), they will try to move to the first step of stair-2 (provided it does not create an overlap), with a preferential speed and step length still conforming to the platform rules.

## 4. Validation of the model and results

We now proceed to testing and validating the EOSM model by comparing its output to the data that we extracted from our passive empirical observations. To this end, some characteristics of the pedestrian flow are empirically measured and used as input in the model:

(1) On the staircase, pairs walk down the stairs at a pace randomly drawn from a uniform distribution [1.33 stair/s, 2 stairs/s], whereas on the platform their free walking speed is drawn from a uniform distribution [0.75 m/s, 1m/s].

(2) 10% of pairs are allowed to make detours (detouring attribute).

(3) The simulated injection rates of pairs both at the stair-entrance and at the platform gate are aligned with their empirical values.

Two indicators are adopted in this section to compare the model output with the empirical study: the fundamental diagram (one of the most important characteristics of pedestrian dynamics, as pointed out by Zhang et al. [41]) and the evolution of the number of pedestrians in the measurement area.

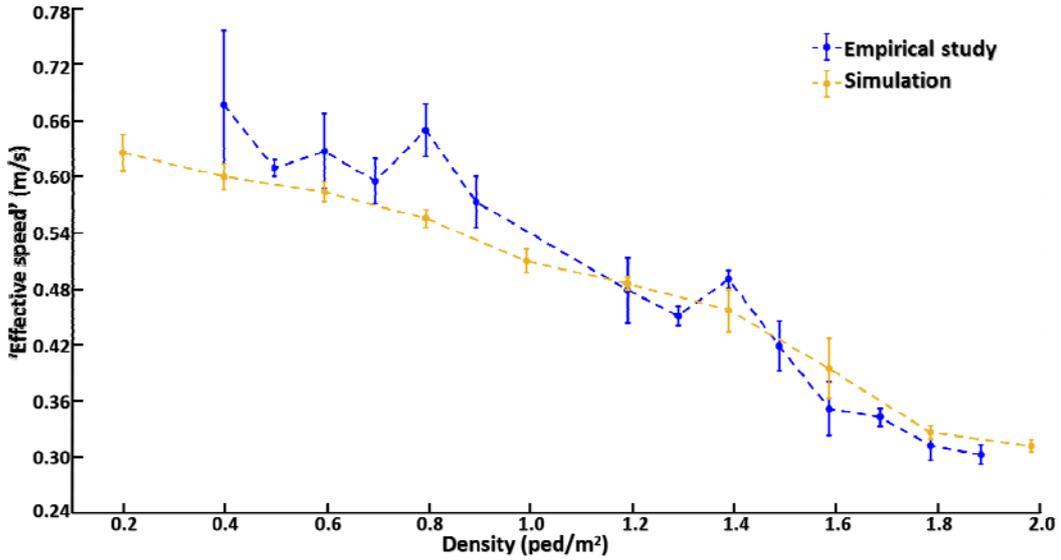

Fig. 9 Fundamental relation between the empirical and numerical 'effective speed' $L/T$ and the density, for pairs coming from the stairs. The bars represent the standard error.

In Fig. 9, we plot the 'fundamental diagram' associated with pairs coming from the stairs based on their movement on stair-1 plus the platform, exclusive of those entering through the platform gate, who have a shorter travel distance through the platform and are less visibly affected by the possible congestion at the merger between the two incoming flows (furthermore, in the context of egress from multi-floor buildings, pedestrians from higher floors deserve more attention, according to Galea et al. and Hakonen et al. [42], as it takes them longer to egress and they face more potential hazards). An effective speed for each pair under consideration is estimated by dividing a typical travel distance $L$ (in practice, $L$=6m approximately, i.e., horizontal walking distance from stair-1 to stair-2) by the measured or simulated travel time $T$ and is plotted as a function of the total density on stair-1 and the platform. Manifestly, the corresponding areas (and trajectories associated therewith) do not lie in a horizontal plane, which results in a conundrum for the definition of areas and speeds. We cut the Gordian Knot by projecting all quantities onto the horizontal plane. As the density increases, the effective speed decreases in a similar way for both the empirical study and the simulation; the agreement between the two curves gets even more satisfactory in the higher-density regime, whereas at low density some discrepancy can be observed, along with larger standard error bars for the empirical study. We rationalize these two points by underlining that, in this low-density regime, pairs have more freedom in choosing their walking speed in reality, following their intentions, unconstrained by the external environment, whereas their speed is preset in the simulations. As the density increases and more interactions take place, the spontaneous choices of people weigh less and less, resulting in a closer match between the simulation and the empirical study.

The 'fundamental diagram' of Fig. 9 reflects the global dependence of the flow on the density, but does not show the accumulation of pedestrians in the facility with time. For that purpose, we present in Fig. 10 the time evolution of the total number of pedestrians on the two staircases and on the platform, evaluated in time intervals of 1 and 5 in the empirical observations and the simulations, respectively. The figure confirms that, under the same input conditions, the changes in the number of pedestrians inside the observation area are very similar. The peak value of the two curves is the same, and the peaks occur at approximately the same time. The different counting time intervals may incidentally explain a considerable part of the remaining difference. Thus, our EOSM model is able to depict the adult-child mixed pedestrian on the staircase and leads to reasonable predictions for the egress process and time, opening the door for an evaluation of egress performance in the presence of adult-child pairs.

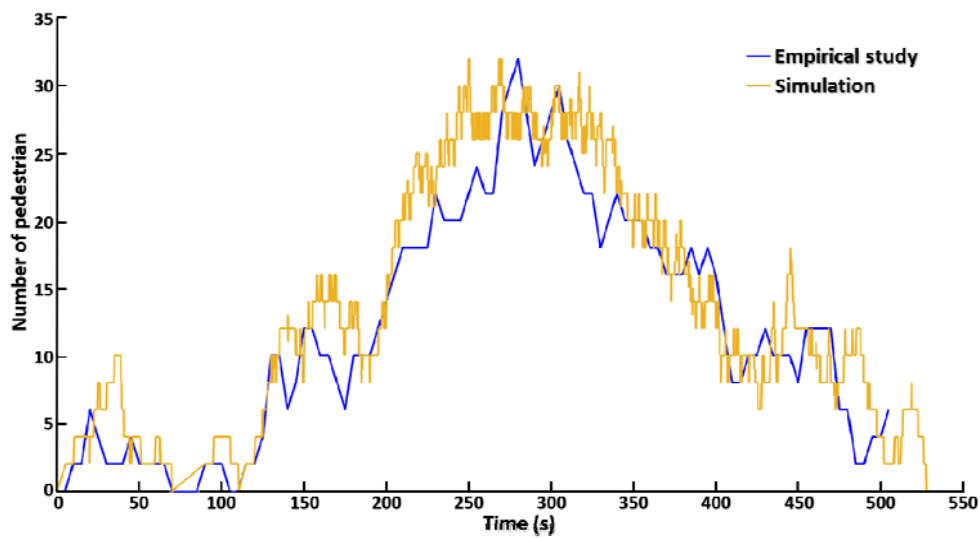

Fig. 10 Time variation of the number of pedestrians inside the measurement area in the empirical study and in the simulation[4].

## 5. Management and guidance strategies

---

[4] An animation of the simulated dynamics is presented in the form of a GIF, as shown in Supplementary Material.

How to enhance the (e.g., emergency) evacuation efficiency of students in a multi-floor primary school, has attracted a great deal of attention (e.g., the research done by Li et al. [5]), and some feasible strategies are proposed to this end (e.g., the grade layout, i.e., students' space allocation arrangement considering grades attributes). Still, crowds consisting of adult-child pairs bring more complexity than crowds consisting of only children. Recalling that the research deals with a training school, we now enquire into possible strategies to enhance the egress efficiency of the adult-child mixed pedestrian flow, here in non-emergency conditions. By enhancing this efficiency, not only can the after-class egress time for the training school be shortened (which is beneficial for the time usage of the premises), but it can also serve as a reference for managing pedestrian emergency evacuation (although it must be conceded that the psychological status of pedestrians may be different in emergency conditions, so that we choose to restrict our attention to managing strategies).

5.1 Effect of the pairing mode

Although three pairing modes were identified in the empirical study, only 10% of pairs opted for the **child:front** mode. From a management perspective, is it worth promoting this pairing mode which occupies less space in the transverse direction? To circumvent the artifacts that may ensue from the unsteady empirical flow, we turn to constant-inflow conditions to address this question numerically: the inflow rates $J_s(pair/min)$ and $J_d(pair/min)$ (averaged over 5-second time windows, i.e., taking every 5-second as the counting unit) of pairs coming from the stairs and the platform-gate, respectively, will be constant for each simulation. Given that crowded settings are more interesting both from a fundamental and from a practical viewpoint, $J_s$ is always set to the maximum inflow value of 48 pairs/min, while $J_d$ will vary from 12 pairs/min to 48 pairs/min, enabling us to study the effect of the pairing mode in situations with little merging flow (low $J_d$) as well as situations with strong interactions with the merging flow (large $J_d$). As for the fundamental diagram, we will focus more on pairs from the higher floor; the fraction of pairs with a detouring capacity remains equal to 10% for all pairs. The average time $T_{1-2}(s)$ spent in the measurement area consisting of stair-1 and platform is an intuitive indicator of the egress efficiency for pairs coming from the stairs; it is presented in Fig. 11 for all considered combinations of inflow rates and variations of pairing modes.

We observe that with the increase of $J_d$, the egress efficiency for pairs from stair-entrance is reduced, which may be largely caused by the merging effect. Besides, irrespective of the value of $J_d$, increasing the **child:front** pairing mode enhances the egress efficiency. The comparative advantage associated therewith gets even more pronounced as the settings get more crowded, at larger $J_d$: the efficiency gap between the cases of 10% and 90% **child:front** pairing modes widens with the increase of $J_d$. To sum up, for managers of training schools or other adult-child gathering facilities, encouraging more **child:front** pairing mode on the staircase might be an efficient strategy to enhance the facility's operating efficiency. Obviously, our numerical study is concerned only with the question of egress efficiency in normal conditions and psychological aspects, as well as experimental tests, should be duly considered for any large-scale implementation of this strategy; our work is a first recognition of such a possibility.

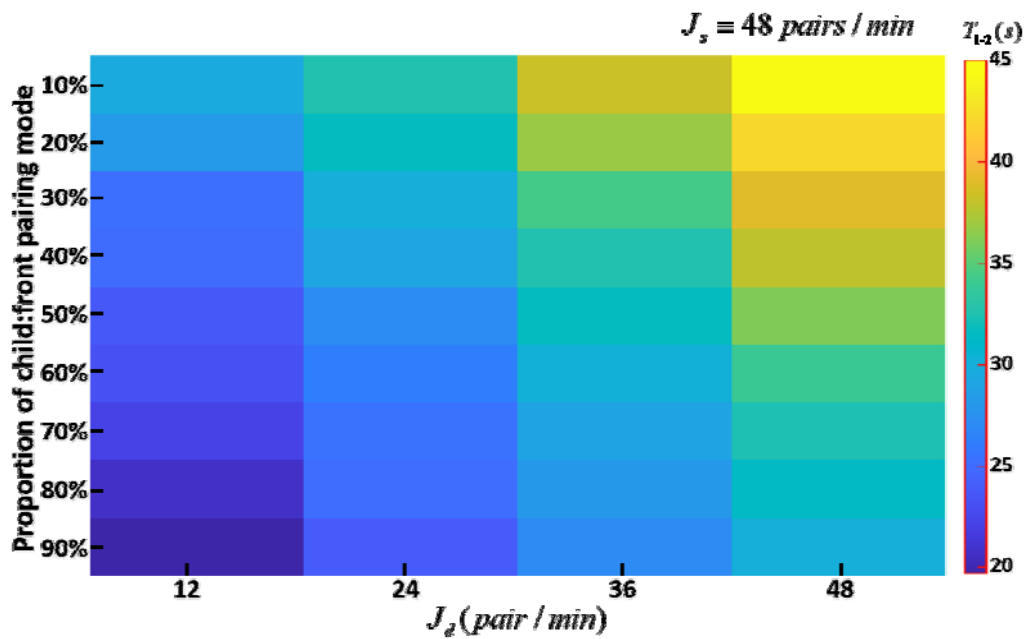

Fig. 11 Time $T_{1-2}(s)$ spent in the measurement area under constant inflow conditions with different **child:front** pairing proportion.

5.2 Effect of the detouring proportion

Some pedestrians always tend to maximize their own benefits in their movement decisions (e.g., bypassing other pedestrians, etc.). As an example, in the context of an evacuation, this 'selfishness' translates into maximizing desired speed in the social force model, which may be detrimental for the collective evacuation time, the 'faster-is-slower' effect coined by Helbing et al. [43]. Other numerical and, more recently, experimental validations of this effect have followed, mainly focusing on crowds of individuals [44, 45].

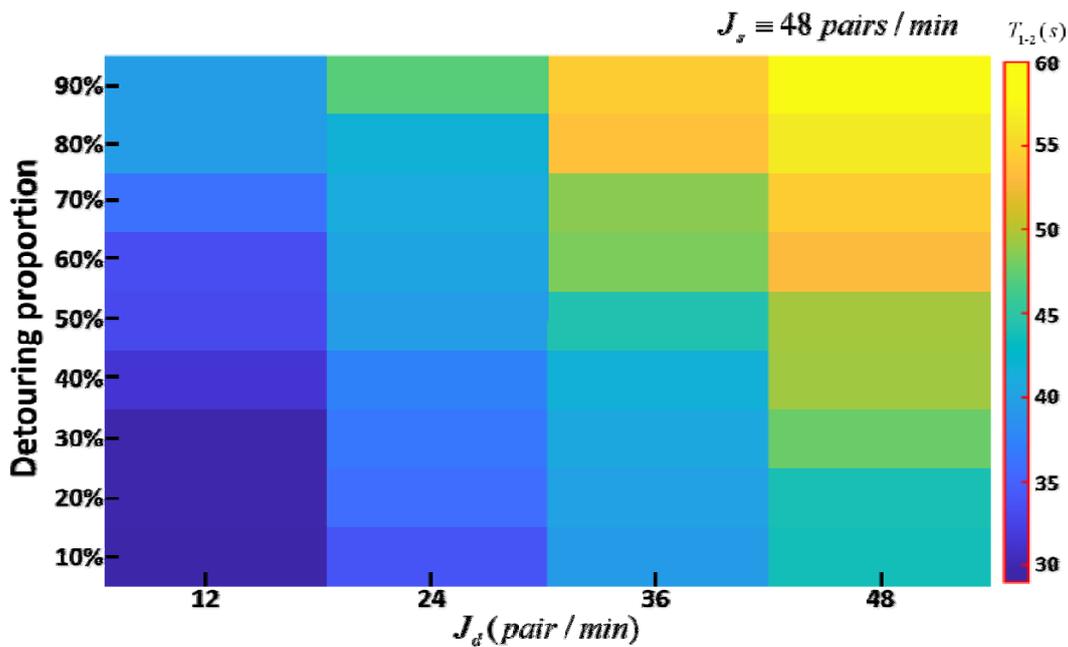

Fig. 12 Time $T_{1-2}(s)$ needed to cross the measurement area, under constant inflow conditions with different proportions of detouring attributes.

Is there an equivalent effect that emerges in the context of our study? If a pair faces obstacles in their intended and original movement route, detouring can help them search for possible positions closer to the target. Hence, the proportion of detouring behavior among all pairs reflects the 'selfish' eagerness to reach the destination as quickly as possible, even if it implies a vying behavior. As in subsection 5.1, we adopt a constant inflow condition and keep the same notations, also focusing on the pairs coming from the stairs. The average time for pairs from the average time $T_{1-2}$ taken by pairs coming from the stairs to cross the measurement area consisting of stair-1 and platform, is also adopted as an indicator of the crowd's egress efficiency. The detailed result is presented in Fig. 12, for a distribution of pairing modes in line with the simulations in Section 4.

Fig. 12 shows that detouring behavior has an adverse effect on the overall egress efficiency for all values of $J_d$, which thus bears some loose similarity with the 'faster-is-slower' mechanism. Otherwise, the tendencies are similar to those already observed in Fig. 11; thus, the figure needs no further explanation. However, it should be stressed that Fig. 12 is only a reflection of our study of a specific scenario with our EOSM. Nonetheless, this simulation, provides inspiration for a strategy to increase the egress efficiency on the staircase, of interest for managers of training schools (or other similar adult-child gathering places); as mentioned above, further validations as to whether and how to apply such strategies based on our numerical findings are out of the scope of this paper.

## 5.3 Effect of inflow desynchronisation

Finally, we turn to a strategy which is less intrusive for the agents, but slightly more demanding in terms of organization. In 2017, commenting on training school's potential safety hazards in the after-class time, the Guangzhou Daily[5] especially underscored that 'during the after-class time, hundreds of pedestrians flood into the staircase, heightening safety hazard and management difficulties' and propounded a management suggestion: shifting the after-class time by 15 minutes on different floors, thus leading to different exit times [46]. However, the extent to which such 'shifts of after-class time' (called inflow desynchronisation below) affect adult-child pedestrian flow on the staircase remains unclear. For sure, if the inflows are fully dephased, pedestrians will spend less time on the staircase. One extreme case is that pedestrians from one specific floor are not allowed to enter the staircase before pedestrians from other floors have left the site; in such case, pedestrians will not go through merging with inflow from other floors. This naturally raises the safety level and the egress efficiency for individuals, but the overall egress efficiency may drop dramatically, because the global egress process is given unlimited time.

---

[5] Guangzhou is one of the biggest cities in China, and the Guangzhou Daily can be regarded as the London Daily to some extent.

A more stringent test would require keeping the global egress time constant. For convenience, we implement this constraint by considering periodic conditions in time. By copying inflows data from entrances and extending them periodically on the time scale, the periodic condition guarantees that, even shifted an arbitrary duration $\tau(min)$, the inflows remain in the time window of interest.

More precisely, the inflow rates and the attributes of pairs coming from the stairs are replicated from the empirical study, but shifted by a duration $\tau\ (min)$ (over a periodic time window) for the pairs coming through the gate (not those coming from the stairs). Besides, the average time $\overline{T_{1-2}}(s)$ spent on stair-1 and platform by pedestrians coming from the stairs remains our main indicator of egress efficiency. The detailed results are shown in Fig. 13. The maximum time shift $\tau$ for the first phase is 8 minutes since the record time duration for the empirical study is 505 seconds; in practice, for $\tau = 9, 18, 27\ (min)$, inputs are set as the same to $\tau = 0\ (min)$ (i.e., the initial case). Phase number is given a symbol as $k\ (k = 1, 2, 3, 4)$. Given inputs are extended periodically, $\overline{T_{1-2}}(s)$ also appears to be periodic with $\tau = 9\ (min)$ as duration.

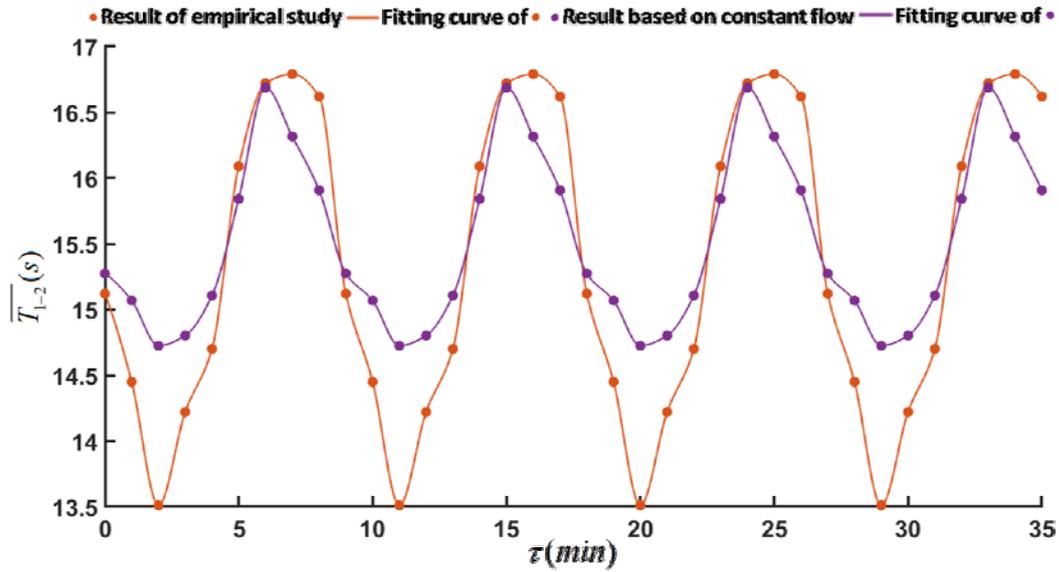

Fig. 13 Average time $\overline{T_{1-2}}(s)$ spent in the measurement area if the egress from the higher floor is time-shifted by a delay $\tau$ (*min*). Orange curve: bona fide simulation using the evolution of inflow rates. Purple: reconstructed chronogram using constant flow simulations with different $\tau$ (*min*).

Fig. 13 establishes the success of inflow desynchronisation in improving the crowd's egress efficiency. The crowd's egress efficiency reaches the best performance when $\tau = 2 + 9 \cdot (k-1)$, whereas $\tau = 7 + 9 \cdot (k-1)$ for simulation gives the worst egress efficiency. From a management's point of view, what matters most is the $k = 1$ case since the shifting of inflows should always consider the overall facilities' operating efficiency; thus, considering the empirical study case, by desynchronizing inflows by only 2 minutes, the egress efficiency may be improved reasonably. Nevertheless, for practical applicability, it would rather be advisable to set the shifting delay case by case, without a univocal standard and universal solution; this flexibility is all the more needed as some delays $\tau$ can have an adverse effect on the egress efficiency.

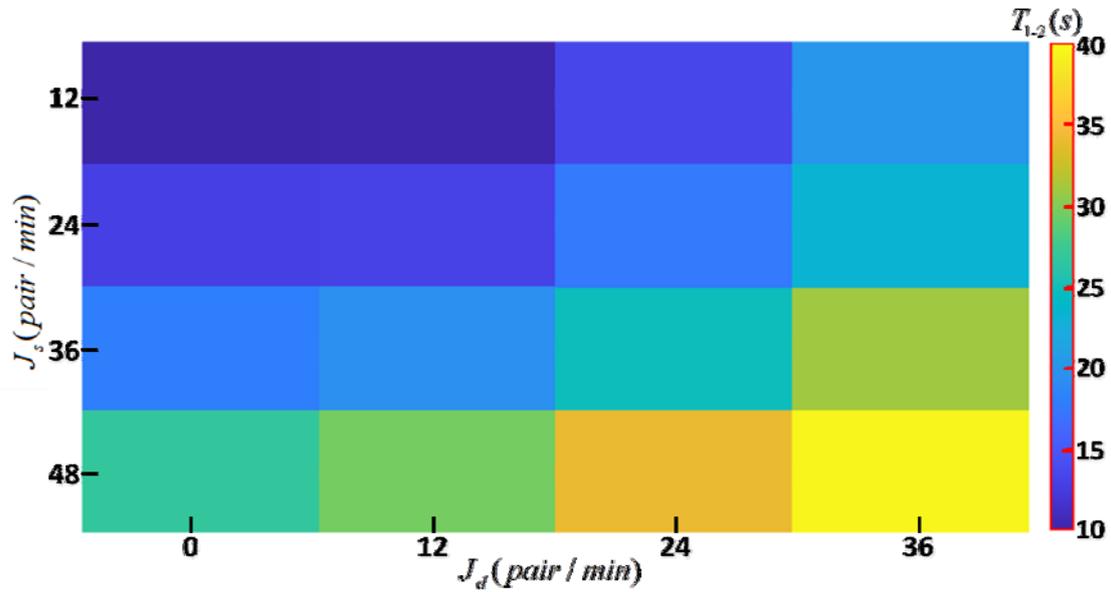

Fig. 14 Time $T_{1-2}(s)$ needed to cross the measurement area for each pair $(J_s^i, J_d^i)$ [6].

To gain deeper insight into these findings and, in particular, into the effect of the variations in time of each inflow, we try to replicate the results of Fig. 13 (where time-varying inflow rates are used, in line with the empirical observations) by using and combining only simulations with constant inflow rates. More precisely, we run a series of simulations with constant inflow rates $(J_s^i, J_d^i)$ from the stairs and from the gate and measure the average time $T_{1-2}^{(J_s^i, J_d^i)}$ spent in the measurement area for each $(J_s^i, J_d^i)$ pair (see Fig. 14). Then, for each shifting delay $\tau$, we assess how many pairs walked in when the inflow rates were around $(J_s^i, J_d^i)$ and approximate $\overline{T_{1-2}}$ under the desynchronisation strategy (with varying inflows) as the average of $T_{1-2}^{(J_s^i, J_d^i)}$ weighted by the foregoing number of pedestrians, viz. $\overline{T_{1-2}} \sim \sum_{i=1}^{N_t}(P_i^s \cdot T_{1-2}^{(J_s^i, J_d^i)})/P^s$, where $N_t$ denotes the total number of time intervals; $P_i^s$ denotes the number of pedestrians coming from the stairs in the i-th time interval; $P^s$ denotes the total number of pedestrians coming from the stairs, from 1 to $N_t$. The resulting estimates based on constant-flow intervals, shown in purple in Fig. 13, display the same trends with $\tau$ as the bona fide simulations. In particular, the travel times $\overline{T_{1-2}}$ peak at approximately the same value. This confirms that the desynchronisation effect is not an artifact of our simulations with variable inflows, but rather results from the interactions in the crowd. Beyond this qualitative agreement, discrepancies are however visible, especially in the situations of highest-efficiency, where constant-flow recombination approximations appear to underestimate the possible gains in efficiency, possibly because congestion is less severe under temporary peak flows than where these peak flows persistent (as they are in the constant-flow simulations); this highlights the effect of the shape of the curve of evolution of the 'demand' in time, beyond the series of instantaneous values that it takes.

---

[6] In the empirical data, the maximum input rate from the gate is 36 pair/min. Thus, when simulating the constant flow results for testing inflow desynchronisation, cases with 48 pair/min are not included.

## 6. Summary and discussion

In this paper, we studied adult-child mixed pedestrian flows on staircases, motivated by the still inchoate knowledge about heterogeneous crowds and the fact that staircases are an essential element in multi-floor buildings, where stampedes may occur. Field data in such a scenario were first collected and analyzed, revealing a prevalence of side-by-side pairs, but also **child:front** pairs, the semi-circular path along which pairs move on the platform, led by the outer pedestrian, a decrease in the flow rate and an increase in the number of steps to cross the platform as the density gets higher, and a low propensity overall to overtake other pairs by making a detour when there is congestion. Based on these observations, we designed a temporally discrete and spatially continuous pedestrian flow model (EOSM), which extends the initial Optimal Step Model by incorporating pairs of agents, variable step-length choices, and introducing more realistic (elliptic) pedestrian shapes. The ability of the EOSM to reproduce the observed collective dynamics was validated quantitatively, notably by examining the fundamental diagram, the time evolution of the number of people in the measurement area, and the time spent on the platform. Thus validated, the model provided a flexible tool to test and assess management and guidance strategies aimed at improving the egress efficiency, and further being an reference for improvement of emergency evacuation performance. In this regard, encouraging **child:front** pairing may be beneficial from an operational viewpoint, whereas increasing the propensity to make detours leads to a reduced overall efficiency. The least intrusive strategy, as far as individuals are concerned, consists in desynchronizing the inflows of children and adults, by shifting the ends of classes at the different floors. Even if the global time window during which the egress takes place is kept fixed, we found that introducing such a delay could have a significant positive impact on the egress efficiency; in practice, the best shifting delay should be carefully adjusted case by case (bearing in mind that poorly chosen delays may worsen the flow conditions).

To conclude, we would like to mention some limitations and prospective improvements of this work. First of all, owing to videotaping limitations, the data could only be counted manually and we did not have access to the detailed pedestrian trajectories. Secondly, the present implementation of the EOSM only considers adult-child pairs, exclusive of children or adults walking alone (or other group compositions), although these may be rare but also be encountered at training schools; it is thus oblivious to the behavior of single pedestrians and their interaction with pairs. Last but not least, although the management and guidance strategies that we put forward were tested using a tool that we took pains to validate in the scenario under study, the empirical verification should be conducted before one can consider widely enforcing these strategies in practice, to consolidate their effectiveness and check for overlooked side effects. These are natural next steps in the wake of this first work.

## Acknowledgement

(1) This work is supported by: i) National Natural Science Foundation of China (71771005, 72171006); ii) Agence Nationale de la Recherche under project name MADRAS (ANR-20-CE92-0033); iii) Institut Rhônalpin des Systèmes Complexes (IXXI); iv) China Scholarship Council (202006020212).

(2) The authors wish to thank Mr. Iñaki Echeverría-Huarte (Ph.D. student at Universidad de Navarra, Spain) for his technical advice regarding simulation and Mr. Si-Qi Chen (undergraduate at Beihang University, China) for his assistance of video-data analyzing.

## Authorship contribution statement

Chuan-Zhi Xie: Conceptualization, Methodology, Software, Investigation, Experiment, Data analysis, Writing-original draft. Tie-Qiao Tang: Conceptualization, Experiment, Supervision. Bo-Tao Zhang: Investigation, Experiment, Proofreading. Alexandre Nicolas: Conceptualization, Methodology, Software, Data analysis, Writing-polishing, Supervision.